\documentstyle[PASJadd]{PASJ95}
\draft
\def\stackls{\raisebox{-1ex} {$\; \stackrel{\textstyle <}{\sim}\;$}}
\def\stackrs{\raisebox{-1ex} {$\; \stackrel{\textstyle >}{\sim}\;$}}
\def\be{\begin{equation}}
\def\ee{\end{equation}}
\def\ni{\noindent}
\def\vs{\vspace}
\def\hs{\hspace}

\begin{document}

\title{Analytical Studies on the Structure and Emission \\ of the SS 433 Jets}
\author{
Hajime {\sc INOUE, }$^{1}$
Noriaki {\sc SHIBAZAKI, }$^{2}$
and
Reiun {\sc HOSHI}$^{2}$
\\[12pt]
$^{1}$ {\it The Institute of Space and Astronautical Science, 3-1-1
Yoshinodai, Sagamihara, Kanagawa 229-8510} \\ {\it E-mail:
inoue@astro.isas.ac.jp} \\
$^{2}$ {\it Department of Physics, Rikkyo University, 3-34-1
Nishi-Ikebukuro, Tokyo 171-8501} }

\abst{We study the structure and emission of the SS 433 jets in the X-ray
emitting region and in the inner and hotter portion inside the X-ray
emitting region. In order to consider the jet structure from the inner to
outer regions we develop the hybrid model combining the conical beam and the
model beam whose cross section grows with the distance more slowly. We find
that the jet beams in the inner and hotter portion are of two-temperature
and emit a large amount of high energy gamma photons. Our analyses suggest
the thick absorbing envelope to exist in the SS 433 system. Based on our
results, we discuss the possible acceleration mechanism for the SS 433 jets.
}

\kword{Astrophysical jets --- Stars: individual (SS 433) --- X-rays: binaries}

\maketitle
\thispagestyle{headings}

\begin{center}
\large{(Accepted for publication by Publ. Astron. Soc. Japan)}
\end{center}

\section{Introduction}

The galactic source SS 433 is well-known as an unique object exhibiting a
jet phenomenon ( for review, see Margon 1984 and Vermeulen 1989). The moving
behavior of Balmer and He I lines has been interpreted successfully in terms
of the two opposing jets precessing with a period of 164 days. The velocity
of matter outflowing in the jet is found to be 0.26$c$ and remarkably
stable, where $c$ is the speed of light. The SS 433 system is an eclipsing
binary with the orbital period of 13 days. The system may consist of an
early type star and an accretion disk around a compact object. The compact
star may be a neutron star or a black hole, although the definite
determination has been still waited.

SS 433 has attracted huge attention from the time of the discovery because
of its peculiar behavior. A lot of observations have been conducted in the
radio, optical and X-ray bands. Many theoretical studies have also been put
forward. In spite of these enormous efforts the most fundamental problems on
SS 433 still remain unsolved. Those are the nature of the compact star and
the acceleration and collimation of the jet beams.

Recently, the ASCA satellite has detected X-ray lines of various elements,
such as Fe, Ni, Mg, Si, S and Ar, from the SS 433 jets ( Kotani et al. 1996;
Kotani et al. 1997). In the Fe case both the blue-shifted and red-shifted
X-ray lines, which are emitted respectively from the approaching and
receding jets, have been observed at the same time. Fitting the analytical
jet model to these ASCA data, Kotani et al. (1996, 1997) have examined the
properties of the X-ray emitting portion of the jet beams. They have derived
the important constraints on the jet parameters such as the density,
temperature and size of the X-ray emitting region and the mass outflow rate.
Their analyses have also implied the X-ray absorbing gas to exist in the
system.

We study the structure and emission of jet beams in the X-ray emitting
region and in the inner and hotter portion inside the X-ray emitting region.
Based on the results on the jet structure, we discuss the acceleration
mechanism for the SS 433 jets.

We present a simplified model for the SS 433 jets in section 2. We derive
the analytic solutions for the structure and emission of the X-ray jets in
section 3. Deriving the analytic solutions, we show that the jet structure
is of two-temperature in the inner and hotter portion inside the X-ray
emitting region in section 4. In the final section we discuss the
appropriate jet model, the acceleration mechanism and the absorbing envelope
based on our analytical results on the jet structure and emission.

\section{Model}

The jet may be divided into several characteristic regions depending on the
distance from the central engine, the inner region, the X-ray emitting
region, the optical and radio emission regions, and the outer region. Here,
we focus on the X-ray emitting region and the hotter region inside of it.

No observational evidence is found for the velocity variation along the jets
in SS 433. This fact implies that the matter entrainment from the ambient
medium and other braking processes are unimportant in the region of our
interest. Hence, we assume the matter velocity constant, $v = 0.26c$, along
the jets.

The mass outflow rate $\dot{M}$ in a jet is written as

\be
\dot{M} =S \rho v \quad ,
\ee
\vs{2mm}

\ni where $r$ is the distance from the central engine, $\rho$ the density
and $S$ the area of the jet cross section. For simplicity, we consider the
beam shapes as expressed by

\be
S = \pi R^2 \propto r^n \quad ,
\ee
\vs{2mm}

\ni where $R$ is the radius of the jet cross section and $n$ the constant
number (Fukue 1987). In the following we examine the jet structure and
emission especially for the cases of $n = 1$ and $n = 2$. The case of $n =
2$ corresponds to the conical beam, while the case of $n = 1$ to the jet
beam whose cross section grows with the distance more slowly than that of
the conical beam.

The matter in the jets cools, expanding and emitting radiations. Writing the
temperature and radiative loss rate respectively as $T$ and $\Lambda$, we
have for the energy equation

\be
\frac{3}{2} \; \frac{k}{\mu m_H} \; \frac{dT}{dr} - \frac{kT}{\mu m_H} \;
\frac{1}{\rho} \; \frac{d\rho}{dr} = - \; \frac{\Lambda}{v} \quad , \ee
\vs{2mm}

\ni where $m_H$ is the mass of hydrogen, $k$ the Boltzmann constant and
$\mu$ the mean molecular weight of matter. The second term on the left hand
side of equation (3) and the term on the right hand side represent the
adiabatic and radiative coolings, respectively.

We normalize physical quantities with those at a fiducial point $r_0\; ; \;
x = r/r_0$, $y = T / T_0$, $z= \rho / \rho_0$ and $f = \Lambda/ \Lambda_0$.
Here and in the following the subscript 0 is used to denote the physical
quantities at $r = r_0$. Using these parameters, equations (1) and (3)
reduce to the dimensionless equations,

\be
z = x^{-n}
\ee

\ni and

\be
\frac{dy}{dx} + \frac{2}{3} \; n \frac{1}{x} \; y = - \xi_0 f \quad , \ee
\vs{2mm}

\ni respectively. In equation (5) the free parameter $\xi_0$ is the ratio of
the flow time $t_{f0}$ to the radiative cooling time $t_{r0}$,

\be
\xi_0 = \frac{t_{f0}}{t_{r0}} \quad ,
\ee
\vs{2mm}

\ni where $t_{f0}$ and $t_{r0}$ are defined respectively by

\be
t_{f0} = \frac{r_0}{v}
\ee

\ni and

\be
t_{r0} = \frac{\frac{3}{2} \; \frac{kT_0}{\mu m_H}}{\Lambda_0} \quad . \ee
\vs{2mm}

\ni Note that the flow time also expresses the adiabatic cooling time due to
the plasma expansion. Equation (5) shows that the free parameter $\xi_0$
determines the property of solutions and hence the thermal structure of the
jet.

\section{One-Temperature Jets}
We consider here the region of the jet beams, where a typical temperature is
$\sim$ 10 keV and X-rays are emitted. We assume that the jet is optically
thin to X-ray photons, satisfying $\tau_e = \kappa_e \rho R \stackls 1$,
where $\tau_e $ is the optical thickness and $\kappa_e$ is the electron
scattering opacity. In this region the ion-electron scattering occurs so
frequently that the electrons and ions have the same temperature.

\subsection{Analytic Solutions}
When the radiative cooling rate is given by the thermal Bremsstrahlung
emission,

\be
\Lambda = \Lambda_{f\!f} = 5.7 \times 10^{20} \rho \sqrt{T} \quad {\rm
ergs\; g}^{-1} {\rm s}^{-1} \quad , \ee
\vs{2mm}

\ni the normalized cooling rate is obtained as

\be
f = z \sqrt{y} \quad .
\ee
\vs{2mm}

\ni Equation (5) together with equations (4) and (10) can be solved
analytically to yield

\be
\sqrt{y} = \left \{
\begin{array}{ll}
\frac{1}{x^{1/3}} + \frac{3}{2} \; \xi_0 \left( \frac{1}{x^{1/3}} - 1
\right) & {\rm for} \; \; n = 1 \\[3mm] \frac{1}{x^{2/3}} + \frac{3}{2} \;
\xi_0 \left( \frac{1}{x} - \frac{1}{x^{2/3}} \right) & {\rm for} \; \; n = 2
\quad . \end{array}
\right.
\ee
\vs{2mm}

\ni Note that the solutions (11) tend to the adiabatic cooling formula, $y =
1/x^{2/3}$ and $y = 1/x^{4/3}$ respectively for $n = 1$ and $n = 2$, as
$\xi_0$ approaches to zero. If we intergrate the Bremsstrahlung emission
along the jet, we have for the X-ray luminosity emitted from the region
outside the distance $x$,

\be
L(x) = S_0 r_0 \rho_0 \Lambda_{f\!f0} \; g(x) \quad {\rm ergs \; s}^{-1}
\quad , \ee

\ni where

\be
g(x) = \left \{
\begin{array}{ll}
3 \left( 1 + \frac{3}{2} \; \xi_0\right) \left( \frac{1}{x^{1/3}} -
\frac{1}{x_{\rm max}^{1/3}} \right) - \frac{3}{2} \; \xi_0 \ln \left(
\frac{x_{\rm max}}{x} \right) & {\rm for} \; \; n = 1 \\[3mm] \frac{3}{5}
\left( 1 - \frac{3}{2} \; \xi_0\right) \left( \frac{1}{x^{5/3}} -
\frac{1}{x_{\rm max}^{5/3}} \right) + \frac{3}{4} \; \xi_0 \left(
\frac{1}{x^2} - \frac{1}{x_{\rm max}^2} \right) & {\rm for} \; \; n = 2
\quad . \end{array}
\right.
\ee
\vs{2mm}

\ni In equation (13) $x_{\rm max}$ denotes the outer end of X-ray emitting
region. As an $x_{\rm max}$ we choose here the distance beyond which the
temperature drops below $\sim$ 0.1 keV.

Adopting the solution for the conical beam $(n = 2)$ at $r / r_0 \geq 1$,
Kotani et al. (1996) have calculated the line X-ray emission from Fe, Ni and
other elements and made the detailed comparison with the ASCA observations
for SS 433. We list the standard set of jet parameters they obtained in
Table 1. In Table 1 we also list the jet parameters derived from using the
solution for the model with $n = 1$. The solution of $n = 1$ also reproduces
well the observed properties, such as the X-ray luminosity ( $\sim 10^{36}$
ergs s$^{-1}$) and temperature ( $\sim$ 20 keV). In both solutions the
fiducial point is chosen to be the base of the X-ray emitting region of the
jet. The length of the X-ray emitting region is obtained as $\ell_X \sim
10^{13}$ cm. The optical thickness across the beam is $\tau_{e0} \sim 0.1$
and hence the jet plasma in the X-ray emitting region is transparent,
justifying the assumption made in the energy equation. Note that the value
of the dimensionless parameter $\xi_0$ is $\sim$ 0.1 or less. This fact
shows that the temperature structure is determined mainly by the adiabatic
cooling loss due to the plasma expansion. We notice that the kinetic energy
$L_K$ carried by the outflowing matter amounts to more than $10^{39}$ ergs
s$^{-1}$ and exceeds the Eddington luminosity for a star of one solar mass.

\subsubsection{Jet inside the X-Ray Emitting Region}

The jet radius $R_0$ of the X-ray emitting region and its distance $r_0$
from the central engine are at least several orders of magnitude larger than
the size of the central engine that drives the jet, a neutron star or a
black hole. Although Kotani et al. (1996, 1997) have truncated the inner
portion of the jet, it is quite natural to assume that the jet beam extends
more down to the vicinity of the central engine. In the following let us
examine the property of the jet expected to exist inside the X-ray emitting
region.

We apply the solution (11) to the inner region adjacent to the X-ray
emitting region. Following equations (4) and (11), the density and
temperature of the jet increase with decreasing distance. Hard X and soft
gamma rays are emitted from the jet. The energy transfer from ions to
electrons through the coulomb scattering becomes less efficient at higher
temperatures, whereas electrons lose energy promptly radiating X and gamma
rays and expanding adiabatically. The insufficient energy transfer makes the
ion and electron temperatures different at $T \stackrs 10^9$ K (see the next
section for detail). The jet reaches the critical temperature $T_1 \sim
10^9$ K at the distance $r_1$. Here and in the following we denote the
physical parameters at the innermost region of the one-temperature jet by
the subscript 1. We list the physical quantities at $r = r_1$ in Table 2.
The relativistic Bremsstrahlung emission rate and adiabatic cooling rate,
which are mentioned in the next section, are used to calculate $\xi_1$ in
Table 2. The optical thickness across the beam is less than one, although
approaches one, and the jet is still optically thin against electron
scattering near $r = r_1$. Note that the parameter $\xi_1$ in the case of $n
= 2$ (conical case) exceeds, though a little, one. The radiative loss as
well as the adiabatic loss also contribute significantly to the plasma
cooling near the innermost region of the one-temperature jet in the conical
case. In the $n = 1$ case the cooling of the jet plasma is still determined
by the adiabatic loss. The luminosity of X and soft gamma rays emitted from
the region at $r \geq r_1$ is calculated from equations (12) and (13) and is
listed in Table 2. In the $n = $2 case the luminosity of soft gamma rays
emitted from the region between the distances $r_1$ and $r_0$ exceeds the
luminosity of X-rays ($\sim 10^{36}$ ergs s$^{-1}$) emitted from the region
outside the distance $r_0$ approximately by an order of magnitude, while in
the $n = 1$ case the both luminosities are comparable.

\section{Two-Temperature Jets}

We have the two-temperature jet inside the distance $r_1$, where the ion
temperature is higher than the electron temperature. Ions transfer energy to
electrons through the ion-electron scattering, which in turn lose energy by
the radiative emission and adiabatic expansion. The distribution of ion
temperature along the jet beams can also be described by equation (3),
taking $\mu = 1$ and substituting the ion energy loss rate $\Lambda _{ie}$
for the radiative loss rate $\Lambda$. The ion energy loss rate $\Lambda
_{ie}$ is given approximately by

\be
\Lambda = \Lambda_{ie} = \frac{3}{2} \; \frac{kT}{m_H} \; \nu_E \quad , \ee
\vs{2mm}

\ni where $T$ now expresses the ion temperature and $\nu_E$ is the
ion-electron collision frequency. The ion-electron collision frequency is
written as

\be
\nu_E = 3.6 \times 10^{22} \frac{\rho}{T_e ^{3/2}} \quad {\rm s}^{-1} \quad
, \ee
\vs{2mm}

\ni where $T_e$ is the electron temperature.

We adopt the relativistic Bremsstrahlung emission with the Compton
amplification for the radiative energy loss of electrons and approximate the
emissivity by

\be
\Lambda_{f\!f} = 4.9 \times 10^{16} \eta \rho T_e \quad {\rm ergs \; g}^{-1}
{\rm s}^{-1} \quad , \ee
\vs{2mm}

\ni where $\eta$ is the Compton amplification factor. Equation (16) can be
used approximately even in the optically thick case if the photon diffusion
time across the jet is shorter than the flow time. Energy loss rate of
electrons due to the adiabatic expansion can be approximated by

\be
\Lambda_{ad} = \frac{k T_e}{m_H} \; \frac{1}{t_f} \quad . \ee
\vs{2mm}

\ni The electron temperature is determined from the balance between the
energy acquired from ions and the energy lost to radiation and adiabatic
expansion. When the radiative loss is dominant over the expansion loss
($\Lambda_{f\!f} \gg \Lambda_{ad}$), the electron temperature is expressed
as

\be
T_e = 3.8 \times 10^5 \eta^{-2/5} T^{2/5} \quad{\rm K} \quad , \ee
\vs{2mm}

\ni whereas when $\Lambda_{f\!f} \ll \Lambda_{ad}$,

\be
T_e = 2.0 \times 10^5 T^{2/5} \quad{\rm K} \quad . \ee
\vs{2mm}

\ni In equation (19) the physical parameters for the $n = 1$ case in Table 1
are used. If we equate the ion and electron temperatures in equations (18)
and (19), we can derive the critical temperature $T_1$ above which the jet
plasma is in the two-temperature regime,

\be
T_1 \sim 2 \times 10^9 \eta^{-2/3} \; \; {\rm K} \hs{8mm} {\rm for} \; \;
\Lambda_{f\!f}\gg \Lambda_{ad} \ee
\vs{2mm}

\ni and

\be
T_1 \sim 6.9 \times 10^8 \; \; {\rm K} \hs{8mm} {\rm for} \; \;
\Lambda_{f\!f}\ll \Lambda_{ad} \quad . \ee
\vs{2mm}

We choose the position of $r \sim r_1$ as the new fiducial point and
normalize the physical quantities with those at $r = r_1$ ; $x = r / r_1$ ,
$y = T / T_1$ , and $z = \rho / \rho_1$. The dimensionless equation (5) with
the free parameter $\xi_1$ replaced for $\xi_0$ is again derived from the
ion energy equation. The parameter $\xi_1$ is now re-defined as the ratio of
the flow time to the ion energy loss time or the ion-electron scattering
time given by

\be
t_{ie} = \nu_E^{-1} \quad .
\ee
\vs{2mm}

\ni Furthermore, the ion energy loss rate is written in the dimensionless
form as

\be
f = zy^{2/5} \quad .
\ee
\vs{2mm}

We solve equation (5) together with equations (4) and (23) and derive the
analytic solution,

\be
y^{3/5} = \left \{
\begin{array}{ll}
\frac{1}{x^{2/5}} + \frac{3}{2} \; \xi_1 \left( \frac{1}{x^{2/5}} - 1
\right) & {\rm for} \; \; n = 1 \\[3mm] \frac{1}{x^{4/5}} + 3 \xi_1 \left(
\frac{1}{x} - \frac{1}{x^{4/5}} \right) & {\rm for} \; \; n = 2 \quad .
\end{array}
\right.
\ee
\vs{2mm}

\ni Using equation (16) and integrating the high energy emission along the
jet, we obtain the luminosity of high energy photons emitted outside the
distance $x$. The luminosity $L(x)$ is approximately given by

\be
L(x) \sim S_ 1r_1 \rho_1 \Lambda_{f\!f1} h(x) \; \; {\rm ergs \; s}^{-1}
\quad , \ee
\vs{2mm}

\ni where

\be
h(x) = \left \{
\begin{array}{ll}
\frac{15}{4} \left( 1 + \frac{3}{2} \; \xi_1 \right)^{2/3}
\frac{1}{x^{4/15}} & {\rm for} \; \; n = 1 \\[3mm] \frac{3}{5} \;
\frac{1}{x^{5/3}} & {\rm for}\; \; n = 2 \quad . \end{array}
\right.
\ee
\vs{2mm}

The outer solutions (11) are connected to the inner solutions (24) at $r =
r_1$, thereby yielding a break in the temperature distribution.  As we
proceed inwards along the jet, the ion and electron temperatures increase
following equations (18), (19) and (24). Moreover, the ion temperature
increases faster than the electron temperature, rendering the jet of
two-temperature. As we move furthermore inwards, we reach the sonic point
where the ion sound velocity becomes equal to the flow velocity. Although
derived by ignoring the pressure gradient force and by assuming the constant
flow velocity, the above solution may be used down to the sonic point as
long as the approximate estimates of the jet parameters are concerned. We
list the physical quantities at the sonic point, which are denoted with the
subscript s, in the Table 3. Here, for simplicity, the compton enhancement
factor is set as $\eta \sim 1$.

In the conical jet case electrons lose energy mainly by emitting radiations.
In Table 3 the radiative cooling is adopted for the $n = 2$ case.  The sonic
point is located very far from the central engine. The ion temperature is
approximately an order of magnitude higher than the electron temperature.
The thermal equilibrium is not complete among ions since the ion-ion
scattering time exceeds the flow time. As seen from the values of $\xi_1$
and $\xi_s$, the radiative cooling as well as the adiabatic cooling
contributes significantly to the energy loss of plasma in the
two-temperature region of the jet. We find that the luminosity of the high
energy photons emitted from the region outside the sonic point amounts to
$\sim 2 \times 10^{39}$ ergs s$^{-1}$, which is approximately comparable to
the kinetic luminosity of the outflowing matter in the jet beam.

In the $n = 1$ case electrons lose energy mainly by adiabatically expanding.
In Table 3 the adiabatic cooling is adopted for the $n = 1$ case.  The sonic
point is located close to the central engine. As in the $n = 2$ case the
analytic solution also yields the electron temperature an order of magnitude
lower than the ion temperature and large luminosity of high energy photons.
Note, however, that at the sonic point the radius of the jet cross section
$R_s$ is an order of magnitude larger than the distance $r_s$. Although
assumed in the calculation, the uniformity of physical parameters over the
jet cross section may not be guaranteed when $R \gg r$. Hence, the physical
parameters at the sonic point, shown in Table 3, may not be reliable.  We
expect that the analytic solution for the $n = 1$ case can be applicable
down to the point where $R \sim r$. We list the physical quantities there in
the Table 4. The two-temperature solution for the $n = 1$ case can extend
down to the distance $r \sim 2.3 \times 10^9$ cm and cannot reach the sonic
point keeping $R \stackls r$.

Let us consider the hybrid jet model combining the $n = 1$ and $n = $2
cases. We adopt the $n = 1$ case presented above to describe the outer part
of the jet beyond the distance $r \sim 2.3 \times 10^9$ cm, while for the
inner part we adopt the $n = 2$ case (conical jet). The conical jet extends
inwards starting from the boundary with the physical quantities shown in
Table 4. We approximate the electron cooling rate by the radiative loss
rate. The density and ion and electron temperatures increase with decreasing
distance, following equations (4), (24) and (18). We reach the sonic point
at the distance $ r_s \sim 9.2 \times 10^7$ cm and derive physical
quantities such as shown in Table 5. Note that the sonic point in the hybrid
case locates much closer to the central engine compared to the simple
conical case shown in Table 3. The luminosiy of high energy photons emitted
from the region outside the sonic point is also very large and amounts to an
important fraction of the kinetic luminosity of the outflowing matter. The
other properties of the two-temperature jet are more or less similar to
those of the simple conical case.

Fukue (1987) already considered the hybrid model for the SS 433 jets.
Contrary to our hybrid model, he adopted the $n = 1$ and $n = 2$ beams
respectively for the inner and outer parts of jets. The sonic points lie
close to the compact objects similarly to our calculation for the $n = 1$
case (Table 3). The boundary between the $n = 1$ and $n = 2$ beams may
roughly correspond to the base of the X-ray emitting region. If we adopt a
standard set of parameters for the physical quantities at the base (Kotani
et al 1996), we obtain the radius of the jet cross section at the sonic
point which exceeds the distance of the sonic point from the compact stars,
especially in the neutron star case. The uniformity of the physical
parameters, mentioned above, may not be guaranteed also in the hybrid model
by Fukue (1987).

We note that the electron-positron pairs and appropriate compton
amplification in addition to the pressure gradient force should be included
in the calculation in order to have more accurate estimates on the jet
property especially around the sonic point. Even if included, the
qualitative properties may not deviate significantly from those obtained
here.

\section{Discussion and Conclusions}
The jet beams in SS 433 may originate near the compact object, or more
specifically in the inner region of the accretion disk. The X-ray emitting
region of the jet is likely to be located at the distance of $\stackrs
10^{11}$ - $10^{12}$ cm from the compact object. The jet beams must be
accelerated and collimated to reach the terminal velocity within the
distance of $\stackls10^{11}$ - $10^{12}$ cm. The mechanisms of acceleration
and collimation for the jet beams in SS 433 are poorly understood. In
general the gas and radiation pressures and magnetic field are considered as
a possible driving force for the jet acceleration.

If the shape of the SS 433 jets is simply conical in the X-ray emitting
region and also in the inner region adjacent to it, we find that the sonic
point should be located very far from the central and compact object. The
ion temperature at the sonic point is very high and close to the maximum
attained near the compact object. It is quite difficult to keep the jet
matter so hot when the matter outflows from the jet footpoint near the
central object to the distant sonic point. The jet matter will suffer from
the adiabatic cooling and lower the tempetature substantially unless the
shape of the jet is completely columnar in the region from the footpoint to
the sonic point. Let us turn to the jet acceletation in this conical case.
The free fall velocity at the sonic point due to the gravity of the compact
object is more than an order of magnitude smaller than the jet velocity.
Hence, the acceleration due to the gas pressure may not be applicable in
this case. The radiative acceleration is also ruled out since the inner and
hotter portion of the jets is optically thick along the flow direction and
hence photons from outside cannot enter into the jet and impart momentun to
the matter. The magnetic driving still remains as a possible acceleration
mechanism. These arguments indicate that the simple conical model for the SS
433 jets is less likely.

Promising is the hybrid model presented in the previous section. We adopt
the model jet with $n = 1$, where the cross section of the jet grows with
the distance more slowly than that of the conical jet, in order to describe
the outer part of the jet including the X-ray emitting region. We use the
conical jet to describe the inner and hotter part of the jet. This model
explains the properties of observed X-rays well. Furthermore, in the hybrid
model the sonic point can come close to the central object. Hence, the gas
pressure as well as the magnetic force may also be considered as a possible
driving force for the jet acceleration. The radiative acceleration is again
ruled out since the inner and hotter portion of the jets is optically thick
along the flow direction. We note that the inner and hotter portion of the
jets emit a huge amount of gamma rays. No observational evidence, however,
has not been found for an emission of such high energy photons. Hence, we
conclude that the inner and hotter portions of the jets are hidden by the
surrounding matter, most probably by the thick envelope and accretion disk.
Note that such an envelope, or a dense outflowing atmosphere which envelops
the system, is also suggested from the substantial fluctuations of the
optical magnitudes and the infrared spectrum (Vermeulen 1989; Band 1987).

The collimation of the jet beams is another important issue in the study of
the astrophysical jet. The hybrid model above indicates that the jet beam
may be squeezed and collimated more strongly around the distance of $\sim
10^{9}$ cm. We expect that at squeezing a part of the kinetic energy of the
jet matter may be dissipated into heat and reheat the jet matter. A
substantial amount of surrounding matter is also required, again, if it is
responsible for squeezing, from the pressure balance with the ram pressure
of the jet matter. If the thick accretion disk channels the jet beam and
precesses, the precession of the jet would be a natural consequence. The
magnetic confinement is another possibility for collimation. Our studies
show that the magnetic fields, which yield the magnetic pressure comparable
to the gas pressure at the sonic point, are of the order of $10^7$ G. This
field strength may also be used to study and set constraints on the
properties of the magnetically driven jets.

We thank N. Kawai and T. Kotani for useful discussions. We also thank the
anonymous referee for useful comments. This research was supported in part
by the Grant-in Aid for Scientific Research (C) (10640234, 12640302).

\section*{References}

\small
\re
Band D. L.\ 1987, PASP 99, 1269
\re
Kotani T., Kawai N., Matsuoka M., Brinkmann W.\ 1996, PASJ 48, 619 \re
Kotani et al.\ 1997, in Proceedings of the Fourth Compton Symposium, Ed C.
D. Dermer, M. S. Strickman, J. D. Kurfess, AIP Conference Proceedings 410,
922 \re
Fukue J.\ 1987, PASJ 39, 679 \re
Margon B.\ 1984, ARA\&A 22, 507
\re
Vermeulen R. C.\ 1989, Ph. D Thesis, University of Leiden

\label{last}

\clearpage

\begin{table*}
\small
\begin{center}
Table~1.\hspace{4pt}Physical quantities at the base of the X-ray emitting
region.\\
\vspace{6pt}

$\begin{array}{ccccccccc}\hline\hline
\rule{0mm}{5mm}n & r_0 & R_0 & \rho_0 & T_0 & \; \; \; \; \; \xi_0 \; \; \;
\; & \!\!\!\!\dot{M} & L(x_0) & L_k \\
  & \; \; (10^{10} {\rm cm}) & (10^{10} {\rm cm} ) & \!\!\!\!(10^{-11} {\rm
g}\; {\rm cm}^{-3}) & (10^8 {\rm K}) &  & (10^{-6} {\rm M}_\odot {\rm
y}^{-1} ) & \!\!\!\!(10^{36} {\rm ergs}\; {\rm s}^{-1}) & \!\!\!\!(10^{39}
{\rm ergs}\; {\rm s}^{-1}) \\[2mm] \hline
\rule{0mm}{5mm}1 & 6.4 &1.2 & 4.1 & 2.3 & 0.05 & 2.3 & 1.0 & 4.4 \\[3mm]
2 & 52 & 2.3 & 1.5 & 2.3 & 0.15 & 3.1 & 0.97 & 5.9
\end{array}$
\end{center}
\vspace{6pt}

\end{table*}

\begin{table*}
\small
\begin{center}
\begin{tabular}{ll}
Table~2.\hspace{4pt}& Physical quantities at the boundary inside of which \\

 & the jet plasma is in the two-temperature regime.
\end{tabular}
\vspace{6pt}

$\begin{array}{ccccccc}\hline\hline
\rule{0mm}{5mm}n & r_1 & R_1 & \rho_1 & T_1 & \; \; \; \; \; \xi_1 \; \; \;
\; & L(x_1) \\
  & \; \; (10^{10} {\rm cm}) & (10^9 {\rm cm} ) & \!\!\!\!(10^{-10} {\rm
g}\; {\rm cm}^{-3}) & (10^8 {\rm K}) &  & \!\!\!\!(10^{36} {\rm ergs}\; {\rm
s}^{-1}) \\[2mm] \hline
\rule{0mm}{5mm}1 & 1.3 &5.5 & 2.0 & 6.9 & 0.67 & 1.9 \\[3mm]
2 & 12 & 5.5 & 2.6 & 20 & 1.6 & 12
\end{array}$
\end{center}
\vspace{6pt}

\end{table*}
\begin{table*}
\small
\begin{center}
Table~3.\hspace{4pt}Physical quantities at the sonic point. \\
\vspace{6pt}

$\begin{array}{cccccccc}\hline\hline
\rule{0mm}{5mm}n & r_s & R_s & \rho_s & T_s & T_{es} & \; \; \; \; \; \xi_s
\; \; & L(x_s) \\
  & \; \; (10^6 {\rm cm}) & (10^8 {\rm cm} ) & \!\!\!\!(10^{-8} {\rm g}\;
{\rm cm}^{-3}) & \!\!\!\!(10^{11} {\rm K}) & (10^9 {\rm K}) &  &
\!\!\!\!(10^{39} {\rm ergs}\; {\rm s}^{-1}) \\[2mm] \hline
\rule{0mm}{5mm}1 & 4.3 &1.0 & 61 & 4.4 & 9.1 & 0.014 & 0.084 \\[3mm]
2 & \; \; \; \; 1.1 \times 10^4 & 5.2 & 2.9 & 4.4 & 17 & 0.64 & 3.8
\end{array}$
\end{center}
\vspace{6pt}

\end{table*}

\begin{table*}
\small
\begin{center}
Table~4.\hspace{4pt}Physical quantities at the distance where $R = r$ in the
$n = 1$ case.
\vspace{6pt}

$\begin{array}{ccccccc}\hline\hline
\rule{0mm}{5mm}r & R & \rho & T &T_e & \; \; \; \; \; \xi \; \; \; \; & L(x) \\
(10^9 {\rm cm}) & (10^9 {\rm cm} ) & \!\!\!\!(10^{-9} {\rm g}\; {\rm
cm}^{-3}) & (10^9 {\rm K}) & (10^9 {\rm K}) & & \!\!\!\!(10^{37} {\rm
ergs}\; {\rm s}^{-1}) \\[2mm] \hline
\rule{0mm}{5mm}2.3 & 2.3 & 1.1 & 4.2 & 1.4 & 0.23 & 1.6
\end{array}$
\end{center}
\vspace{6pt}

\end{table*}
\begin{table*}
\small
\begin{center}
Table~5.\hspace{4pt}Physical quantities at the sonic point in the hybrid
model.
\vspace{6pt}

$\begin{array}{ccccccc}\hline\hline
\rule{0mm}{5mm}r_s & R_s & \rho_s & T_s & T_{es} & \; \; \; \; \; \xi_s \;
\; \; \; & L(x_s) \\
(10^7 {\rm cm}) & (10^7 {\rm cm} ) & \!\!\!\!(10^{-7} {\rm g}\; {\rm
cm}^{-3}) & (10^{11} {\rm K}) & (10^{10} {\rm K}) & & \!\!\!\!(10^{39} {\rm
ergs}\; {\rm s}^{-1}) \\[2mm] \hline
\rule{0mm}{5mm}9.2 & 9.2 & 6.9 & 4.4 & 1.7 & 0.13 & 1.3
\end{array}$
\end{center}
\vspace{6pt}

\end{table*}


\end{document}